\documentclass[sigconf]{acmart}

\newenvironment{itemize-s}%
{\begin{itemize}%
		\setlength{\itemsep}{0pt}%
		\setlength{\parskip}{0pt}}%
	{\end{itemize}}

\usepackage[english]{babel}
\usepackage[utf8]{inputenc}
\usepackage[colorinlistoftodos]{todonotes}
\usepackage{epsfig}
\usepackage[T1]{fontenc} 
\usepackage{listing}
\usepackage{float}
\usepackage{color}
\usepackage{subcaption}
\usepackage{multirow}
\usepackage{balance}
\usepackage{cancel}
\usepackage{adjustbox}
\usepackage{epstopdf}
\usetikzlibrary{positioning,calc}
\usepackage{longtable}
\usepackage{siunitx}
\usepackage{enumerate}
\usepackage{enumitem}
\usepackage{rotating}
\usepackage{hhline}
\usepackage[printonlyused]{acronym}
\usepackage{tabularx}
\usepackage{booktabs}
\usepackage{soul}
\usepackage{xspace}
\usepackage{xurl}

\usepackage{array}
\newcolumntype{x}[1]{>{\centering\arraybackslash\hspace{0pt}}p{#1}}

\acrodef{EPS}{Evolved Packet System}
\acrodef{TTI}{Transmission Time Interval}
\acrodef{UTLA}{Upper Tier Local Authority}
\acrodef{LAD}{Local Authority District}
\acrodef{CA}{Carrier Aggregation}
\acrodef{BSC}{Base Station Controller}
\acrodef{EPC}{Evolved Packet Core}
\acrodef{PDP}{Packet Data Protocol}
\acrodef{GMM}{GPRS Mobility Management}
\acrodef{GGSN}{Gateway GPRS Support Node}
\acrodef{NAS}{Non-Access Stratum}
\acrodef{RAN}{Radio Access Network}
\acrodef{GPRS}{General Packet Radio System}
\acrodef{UTRAN}{UMTS Terrestrial Radio Access Network}
\acrodef{SGSN}{Serving GPRS Support Node}
\acrodef{CN}{Core Network}
\acrodef{RNC}{Radio Network Controller}
\acrodef{MSC}{Mobile Switching Center}
\acrodef{MME}{Mobility Management Entity}
\acrodef{UMTS}{Universal Mobile Telecommunication Systems}
\acrodef{xDR}{eXtended Detail Record}
\acrodef{TAC}{Type Allocation Code}
\acrodef{PoP}{Point of Presence}
\acrodef{IMEI}{International Mobile Equipment Identity}
\acrodef{EMSISDN}{Encrypted Mobile Station International Subscriber Directory Number}
\acrodef{MSISDN}{Mobile Station International Subscriber Directory Number}
\acrodef{MME}{Mobility Management Entity}
\acrodef{VoLTE}{Voice over \ac{LTE}}
\acrodef{MVNO}{Mobile Virtual Network Operators}
\acrodef{CDR}{Call Detail Record}
\acrodef{IMSI}{International Mobile Subscriber Identity}
\acrodef{RAT}{Radio Access Technology}
\acrodef{LPWA}{Low Power Wide Area}
\acrodef{VMNO}{Visited Mobile Network Operator}
\acrodef{HMNO}{Home Mobile Network Operator}
\acrodef{DCH}{Data Clearing House}
\acrodef{GSMA}{\ac{GSM} Association}
\acrodef{GSM}{Global System for Mobile communications}
\acrodef{TAP}{Transferred Account Procedure}
\acrodef{LTE-M}{\ac{LTE} Machine Type Communication}
\acrodef{MTC}{Machine Type Communications}
\acrodef{NB-IoT}{Narrow Band \ac{IoT}}
\acrodef{IOT}{Inter Operator Tariff}
\acrodef{IoT}{Internet of Things}
\acrodef{M2M}{Machine-to-Machine}
\acrodef{FNO}{Fixed Network Operator}
\acrodef{ISP}{Internet Service Providers} 
\acrodef{ASP}{Application Service Providers}
\acrodef{IPX}{IP Packet Exchange}
\acrodef{2G}{Second Generation}
\acrodef{3G}{Third Generation}
\acrodef{4G}{Fourth Generation}
\acrodef{5G}{Firth Generation}
\acrodef{ADB}{Android Debug Bridge}
\acrodef{ASCI}{Advertising Standards Council of India}
\acrodef{ASN}{Autonomous System Number}
\acrodef{CDN}{Content Delivery Network}
\acrodef{DL}{Downlink}
\acrodef{DNS}{Domain Name Service}
\acrodef{EaaS}{Experiment as a Service}
\acrodef{ECDF}{Empirical Cumulative Distribution Function}
\acrodef{HTTP}{Hyper Text Transfer Protocol}
\acrodef{ICMP}{Internet Control Message Protocol}
\acrodef{ISP}{Internet Service Provider}
\acrodef{IXP}{Internet Exchange Point}
\acrodef{LTE}{Long Term Evolution}
\acrodef{MSC}{Message Sequence Chart}
\acrodef{E2E}{End-to-end}
\acrodef{QCI}{QoS Class Identifier}
\acrodef{NDT}{Network Diagnostic Tool}
\acrodef{QoE}{Quality of Experience}
\acrodef{QoS}{Quality of Service}
\acrodef{OS}{Operating System}
\acrodef{APN}{Access Point Name}
\acrodef{RMBT}{RTR Multithreaded Broadband Test}
\acrodef{RTT}{Round-Trip Time}
\acrodef{SIM}{Subscriber Identity Module}
\acrodef{SIMs}{Subscriber Identity Modules}
\acrodef{SLA}{Service-Level Agreement}
\acrodef{TCP}{Transmission Control Protocol}
\acrodef{UDP}{User Datagram Protocol}
\acrodef{UE}{User Equipment}
\acrodef{UL}{Uplink}
\acrodef{NRA}{National Regulatory Authority}
\acrodef{EC}{European Commission}
\acrodef{SGW}{Serving Gateway}
\acrodef{PGW}{Packet Data Network Gateway}
\acrodef{GTP}{GPRS Tunneling Protocol}
\acrodef{MNO}{Mobile Network Operator}
\acrodef{EU}{European Union}
\acrodef{HR}{home-routed roaming}
\acrodef{LBO}{local breakout}
\acrodef{IHBO}{IPX hub breakout}
\acrodef{VoIP}{Voice over IP}
\acrodef{IPG}{Inter-Packet Gap}
\acrodef{KS}{Kolmogorov-Smirnov}
\acrodef{FQDN}{Fully Qualified Domain Name}
\acrodef{MNC}{Mobile Network Code}
\acrodef{MCC}{Mobile Country Code}
\acrodef{MCCMNC}{\ac{MCC}-\ac{MNC}}
\acrodef{SIP}{Session Initiation Protocol}
\acrodef{g}{radius of gyration, (or gyration for short)}
\acrodef{MBB}{Minimum Bounding Box}
\acrodef{BBD}{Minimum Bounding Box Diagonal}
\acrodef{KPI}{Key Performance Indicators}
\acrodef{e}{Entropy}
\acrodef{DL}{downlink}

\newif\ifcomment
\commentfalse

\ifcomment
    \newcounter{AELNumberOfComments}
    \stepcounter{AELNumberOfComments}
    \newcommand{\ael}[1]{\textcolor{purple}{\small \bf [ael\#\arabic{AELNumberOfComments}\stepcounter{AELNumberOfComments}: #1]}}
    \newcounter{DPNOTENumberOfComments}
    \stepcounter{DPNOTENumberOfComments}
     \newcommand{\dpnote}[1]{\textcolor{green}{\small \bf [dp\#\arabic{DPNOTENumberOfComments}\stepcounter{DPNOTENumberOfComments}: #1]}}
     
    \newcounter{MBNumberOfComments}
    \stepcounter{MBNumberOfComments}
     \newcommand{\mb}[1]{\textcolor{red}{\small \bf [mb\#\arabic{MBNumberOfComments}\stepcounter{MBNumberOfComments}: #1]}}
     
     \newcounter{EFNumberOfComments}
    \stepcounter{EFNumberOfComments}
     \newcommand{\ef}[1]{\textcolor{orange}{\small \bf [ef\#\arabic{EFNumberOfComments}\stepcounter{EFNumberOfComments}: #1]}}
    
\else
    \newcommand\dpnote[1]{}
    \newcommand\ael[1]{}
    \newcommand\mb[1]{}
    \newcommand\ef[1]{}
\fi

\copyrightyear{2020} 
\acmYear{2020} 
\setcopyright{acmcopyright}\acmConference[IMC '20]{ACM Internet Measurement Conference}{October 27--29, 2020}{Virtual Event, USA}
\acmBooktitle{ACM Internet Measurement Conference (IMC '20), October 27--29, 2020, Virtual Event, USA}
\acmPrice{15.00}
\acmDOI{10.1145/3419394.3423655}
\acmISBN{978-1-4503-8138-3/20/10}

\hypersetup{draft}

\begin{document}
\title[COVID-19 Impact on MNO]{A Characterization of the COVID-19 Pandemic Impact on a Mobile Network Operator Traffic}
\author{Andra Lutu}
\affiliation{
  \institution{Telefonica Research}
}
\author{Diego Perino}
\affiliation{
  \institution{Telefonica Research}
}
\author{Marcelo Bagnulo}
\affiliation{
  \institution{Universidad Carlos III de Madrid}
}
\author{Enrique Frias-Martinez}
\affiliation{
  \institution{Telefonica Research}
}
\author{Javad Khangosstar}
\affiliation{
  \institution{Telefonica UK}
}

\begin{abstract}

During early 2020, the SARS-CoV-2 virus rapidly spread worldwide, forcing many governments to impose strict lockdown measures to tackle the pandemic. 
This significantly changed people's mobility and habits, subsequently impacting how they use telecommunication networks.
In this paper, we investigate the effects of the COVID-19 emergency on a UK \ac{MNO}. 
We quantify the changes in users' mobility and investigate how this impacted the cellular network usage and performance. 
Our analysis spans from the entire country to specific regions, and geodemographic area clusters. We also provide a detailed analysis for London.  
Our findings bring insights at different geo-temporal granularity on the status of the cellular network, from the decrease in data traffic volume in the cellular network and lower load on the radio network, counterposed to a surge in the conversational voice traffic volume.

\end{abstract}

\maketitle

\section{Introduction}
\label{sec:intro}

After the emergence of SARS-CoV-2 in the province of Wuhan (China) in December 2019, the virus rapidly spread to neighboring countries and to the rest of the world. It was declared by the World Health Organization a Public Health Emergency on January 30th, 2020 and a pandemic on March 11th, 2020 (week 11 of 2020).
As a result, different countries have implemented a variety of interventions to contain the virus.
These included forced or recommended confinement, intended to reduce transmission by reducing contact rates between individuals\cite{kraemer2020effect}.
These policies resulted in a dramatic change in human mobility, which in turn affected the traffic patterns and operations in telecommunication networks.

In this paper, we focus on the cellular network of O2 UK, and evaluate how the changes in people's mobility impacted this \acf{MNO} traffic patterns.
The coronavirus outbreak reached the UK on January 31st 2020, when the first two (imported) cases with the respiratory disease COVID-19 were confirmed in York. 
On March 16th 2020 (week 12 of 2020), the government recommended all citizens to work from home, and on March 20th (also week 12), it implemented the closure of sporting events, schools, restaurants, bars and gyms.\footnote{We indicate the weeks of the year 2020 for the different relevant dates because our analysis in the following sections refers to them in the different graphs. Subsequent weeks also refer to the year 2020, but we omit to mention the year for brevity.}
On the 23rd of March (week 13), the government imposed a lockdown on the whole population, banning all non-essential travel and contact with people outside the home.
London was particularly affected by the outbreak, with 27,000 positive cases at the end of May\cite{GOV1}.

We collect and analyze statistics on the operational status of the \ac{MNO} network, and our main findings are as follow.  

By analyzing cellular network signalling information regarding users' device mobility activity, we observe a decrease of 50\% in mobility (according to different mobility metrics) in the UK during the lockdown period.
We find no correlation between this reduction in mobility and the number of confirmed COVID-19 cases, showing that only the enforced government order was effective in significantly reducing mobility. 
We observe this reduction is more significant in densely populated urban areas than in rural areas. 
We further note regional differences in how people relax the mobility restrictions, with an increase in mobility in London and West Yorkshire in weeks 18-19. 
For London, specifically, we observe that approximately 10\% of the residents temporarily relocated during the lockdown. 

We find that these mobility changes have immediate implications in traffic patterns of the cellular network. 
The downlink data traffic volume aggregated for all bearers (including conversational voice) decreased for all UK by up to 25\% during the lockdown period. 
This correlates with the reduction in mobility we observe country-wide, which results in people likely relying more on the broadband residential Internet access to run download intensive applications such as video streaming. 
We also note a decrease in the radio cell load, with a reduction of approximately 15\% across the UK after the stay-at-home order, which further corroborates the drop in cellular connectivity usage. 
This effect is consistent with the reported surge in traffic for residential ISPs.
This decrease rewound the traffic load on the MNO infrastructure by one year, to levels similar to those of March 2019. 

The total uplink data traffic volume, on the other hand, experienced little changes (between -7\% and +1,5\%) during lockdown. 
This is mainly due to the increase of 4G voice traffic (i.e., VoLTE) across the UK that peaked at 150\% after lockdown compared to the national medial value before the pandemic, thus compensating the decrease in data traffic in the uplink.
At the same time, we observe an increase of more than 100\% in the downlink packet loss error rate for voice traffic on week 10 and 11. 
This was caused by excess of congestion in the interconnection infrastructure \acp{MNO} use to exchange voice traffic, whose capacity was exceeded during the steep surge of voice traffic. 
The error rate reverted its previous levels during the following weeks thanks to rapid response of the network operations.

Finally, we observe mobility changes have different impact on network usage in geodemographic area clusters. In densely populated urban areas, we observe a significantly higher decrease of mobile network usage (i.e., downlink and uplink traffic volumes, radio load and active users) than in rural areas. By looking into the case of London, we observe that this is likely due to geodemographics of the central districts (e.g., Eastern-Central(EC) and Western-Central (WC)), which include many seasonal residents (e.g., tourists), business and commercial areas.

The rest of the paper is organized as follows. In Section \ref{sec:dataset} we describe the measurement infrastructure, as well as the data feeds and the metrics used. In Section \ref{sec:mobility}, we describe the evolution in mobility observed throughout the lockdown imposed in March and April. Then, in Section \ref{sec:performance}, we describe the changes observed in different parameters representative of the MNO's network performance in the UK and specifically for the case of London. Next, we present the related work and conclude the paper.

\section{Dataset}
\label{sec:dataset}

In this section, we describe the measurement infrastructure we leverage for collecting network data from a large commercial MNO in the UK (with more than 25\% market share in the UK in 2019) . 
We detail the dataset we built and the metrics we use to capture the activity of smartphone devices.  


\subsection{Measurement Infrastructure}

\begin{figure}[!t]
	\centering
	\includegraphics[width=\linewidth]{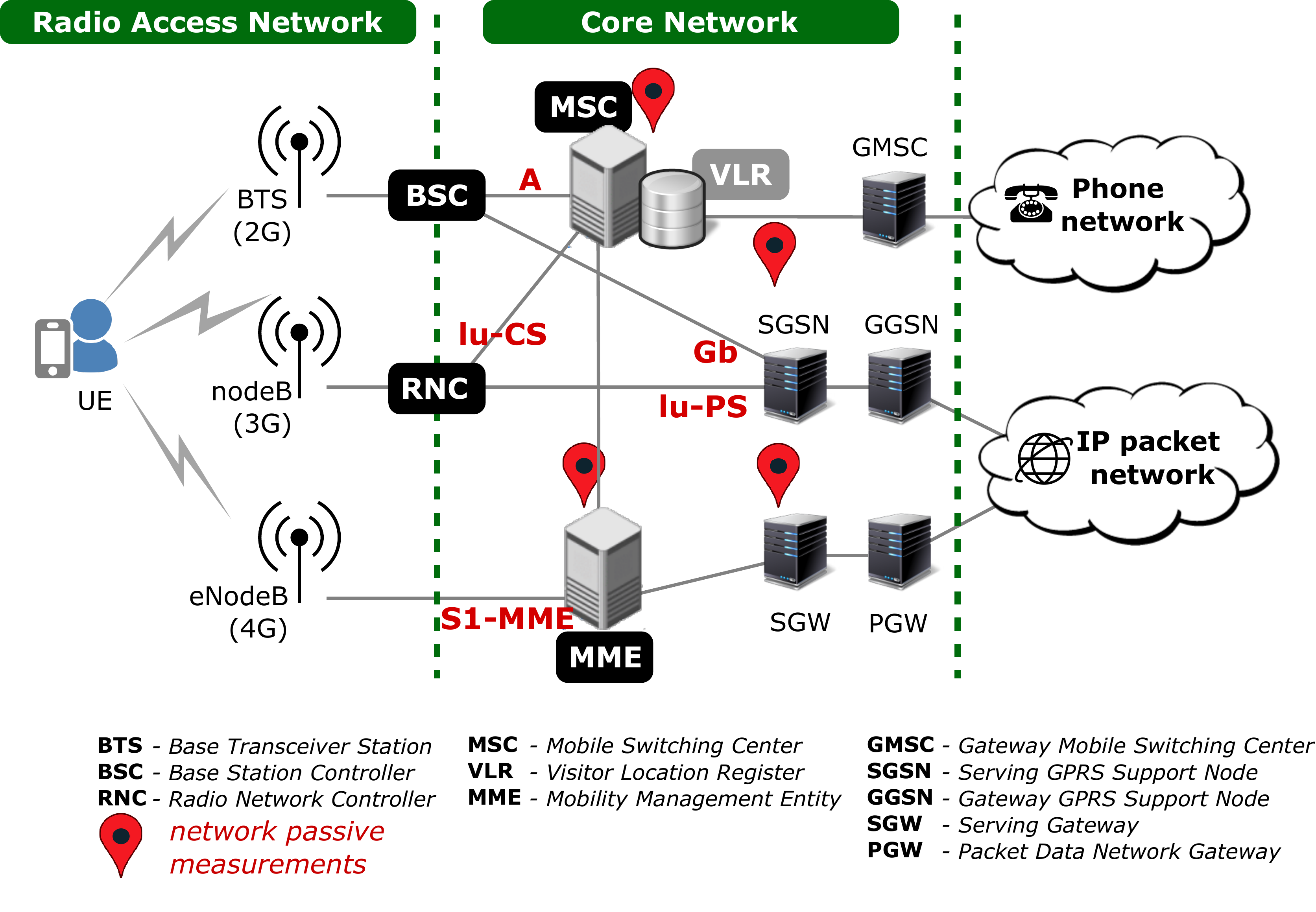}%
	\vspace{-3mm}
	\caption{High-level architecture of the measurement infrastructure integrated in the cellular network. }
	\label{fig:mno_architecture}
	\vspace{-5mm}
\end{figure}

The cellular network we study supports 2G, 3G and 4G mobile communication technologies.  
In Figure~\ref{fig:mno_architecture}, we illustrate a high-level schema of the MNO architecture.
Such a network can be simplified to consist of three main domains: (i) the cellular device (in our case, the smartphone used as primary device by end-users), (ii) the \ac{RAN} and (iii) the \ac{CN}.

Our passive measurement approach relies on commercial solutions the MNO integrates within its infrastructure.
The red pins in Figure~\ref{fig:mno_architecture} mark the network elements that we monitor, namely the \ac{MME}, the \ac{MSC},  the \ac{SGSN}/\ac{SGW}, and the Cell Sites. 
We collect control plane information for both voice and data traffic from the total population of devices connected to the MNO's radio network, as well as \ac{KPI} of cell sites. 

\paragraph*{Cell Sites.}
Cell sites (also called cell towers) are the sites where antennas and equipment of the RAN are placed. Every cell site hosts one or multiple antennas for one or more technologies (i.e., 2G, 3G, 4G), and includes multiple cells and sectors. 
For every cell site we have detailed information including location, radio technologies available, number of cells and radio sectors. 
We collect \ac{KPI} for every radio sectors (e.g., radio load, average user throughput, traffic volume) that we aggregate at postcode level or larger granularity.

\paragraph*{Radio Interfaces.}
We capture and process logs reporting on activities on the lu-PS (for 3G) and Gb (for 2G) interfaces, which carry events related to data packet transmissions and mobility management. 
Similarly, for the LTE networks, we capture the logs at the \ac{MME} nodes on the S1 interface, reporting on mobility management events and bearer management, and user plane S1-UP interface for data and voice (over data) events. 
We also capture and process logs that report on events on the lu-CS (for 3G) and A (for 2G) interfaces, for 2G and 3G voice events. Note that the A interfaces also carries mobility management information. 
For the complete detailed specifications, we direct the reader to ~\cite{3GPP}.


\subsection{Data Feeds}

From our measurement infrastructure, we capture various data feeds from the mobile network that we describe next. Note that these feeds are aggregated at postcode level or larger granularity. 

\paragraph*{General Signaling Dataset.}
As described in the previous section, we capture the activity of the users in the control plane for the different \acp{RAT} supported by the cellular provider. Specifically, for every RAT, the signalling dataset we collect includes the (control plane) signaling messages related to events triggered by the MNO's subscribers, including Attach, Authentication, Session establishment, Dedicated bearer establishment and deletion, Tracking Area Update (TAU), ECM-IDLE mode transition, Service request, Handover and Detach.
Each event we capture carries the anonymized user ID, \ac{SIM} \ac{MCC} and \ac{MNC}, \ac{TAC} (the first 8 digits of the device IMEI, which are statically allocated to device vendors), the radio sector ID handling the communication, timestamp, and event result code (success / failure). 
Further, we aggregate this information at postcode level or larger granularity. 

\paragraph*{Devices Catalog.}
We consider a commercial database provided by \ac{GSMA}. 
This catalog maps the device \ac{TAC} to a set of device properties such as device manufacturer, brand and model name, operating system, radio bands supported, etc.
With this information we are able to distinguish between smartphones (likely used as primary devices by the mobile users) and \ac{M2M} devices. 

\paragraph{Radio Network Topology}
To account for potential structural changes in the radio access network (e.g., new site deployments), we rely on a daily snapshot of the network topology. This includes metadata (location  and configuration) and the status (active/inactive) of each cell tower.  

\paragraph*{Radio Network Performance.}
We rely on a commercial solution the MNO deploys to collect the radio network performance dataset. 
This dataset includes various \ac{KPI}, including average cell throughout, average user throughput, average percentage of resources occupied, average number of users, total volume of data traffic uplink/downlink and total volume of conversational voice traffic. 
We collect this data hourly, and aggregate at postcode level or larger granularity. 

\paragraph{UK Administrative and Geo-demographic Datasets}

We use the National Statistics Postcode Lookup (NSPL) dataset for the UK as at February 2020 to group the postcode areas into \ac{UTLA}. 
The NSPL is produced by ONS Geography, providing geographic support to the Office for National Statistics (ONS) and geographic services used by other organisations.
Furthermore, we use the latest available Area Classification for Output Areas (2011 OAC) released in 2011, which represents a widely used public domain census-only geodemographic classifications in the UK~\cite{gale2016creating}. The 2011 OAC dataset summarizes the social and physical structure of postcode areas using data from the 2011 UK Census, and is updated every 10 years when census is performed (the next one will be available in 2021).


\subsection{Mobility Statistics}
\label{sec:mobility_metrics}


Since we are interested in analyzing mobility of people, we focus on their primary devices. We use the TAC database to filter only the devices that are smartphones (i.e., we drop \ac{M2M} devices such as smart sensors). We are also able to separate the native users of the MNO, and drop the international inbound roamers from further analysis. 
Using the signalling data-set described above, we then associate each (anonymized) user to a radio tower throughout the time they are connected to the MNO's network. 
Based on the radio network topology, we further attach to each radio tower its geographic location (postal code and approximate coordinates). 
With this, we then generate aggregated mobility statistics over six disjoint 4-hour bins of the day (e.g., 04:00AM - 08:00AM, 08:00AM - 12:00PM,  12:00PM -  04:00PM), and also over the entire day (i.e., 24 hours time window). 

For each user, we determine the total duration of time they spend connected to every cell tower and select the top 20 towers.
This allows us to identify all relevant places, as previous studies have shown than more than three quarters of people have between 3 to 6 important places, and in general no more than 8 \cite{isaacman2011identifying, gonzalez2008understanding}.
After doing this initial filtering, we obtain information regarding roughly 22 million native users aggregated at postcode level or larger granularity (e.g., \ac{UTLA} or geodemographic cluster).

\paragraph*{Mobility Metrics}
From the aggregated mobility statistics, we focus on two metrics:  entropy and radius of gyration. 
The combination of both metrics gives a wide view of changes in mobility: while entropy measures the repeatability of movements, radius of gyration is an indication of the distance travelled. 
The two metrics are independent, one could have a high entropy with a reduced gyration, implying someone that moves in a reduced physical space almost randomly; or, on the contrary, have a low entropy with a large value of gyration, implying someone that moves over a large area but repeating the trajectories done.
These metrics are computed over a day for each individual and aggregated to obtain an average value per day.
Even if we compute these metrics per user at cell tower level, we aggregated them at postcode or larger granularity. 

Entropy is a measure of the randomness of the movements of an individual, and as such, a metric for the predictability of movements\cite{song2010limits}. 
From the variety of ways to calculate entropy in mobility \cite{song2010limits}, we implemented a temporal-uncorrelated entropy, that characterizes the heterogeneity of visitation patterns. Formally:

\begin{equation}
e= - \sum_{j=1}^{N} (p(j)\log(p(j))
\end{equation}

with $p(j)$ is the fraction of the time spent in the $\textit{j}^{th}$ visited cell tower (being a proxy for the probability for the user to be in that cell tower). 

Radius of gyration is a key characteristic to model travelled distance \cite{gonzalez2008understanding}, and measures how far from the center
of mass the mass is located~\cite{abramowicz1993concept}. 
It is defined as the root mean squared distance between the set of cell towers and its center of masses. Formally: 

\begin{equation}
g=  \sqrt{\frac{1}{N} \sum_{j=1}^{N} (t_j\textbf{l}_j - \textbf{l}_{cm})^2}
\end{equation}

where $\textbf{l}_j$ represents the location of the $\textit{j}^{th}$ visited cell tower, $t_j$ represents the time spent in the $\textit{j}^{th}$ visited cell tower and $\textbf{l}_{cm}$ represents the location of the center of mass of the user’s trajectory, calculated as $\textbf{l}_{cm} =  \frac{1}{N} \sum_{j=1}^{N} \{ t_j\textbf{l}_j$\} and
$N$ the total number of  towers visited. 


\begin{figure}[!t]
	\centering
	\includegraphics[width=.85\linewidth]{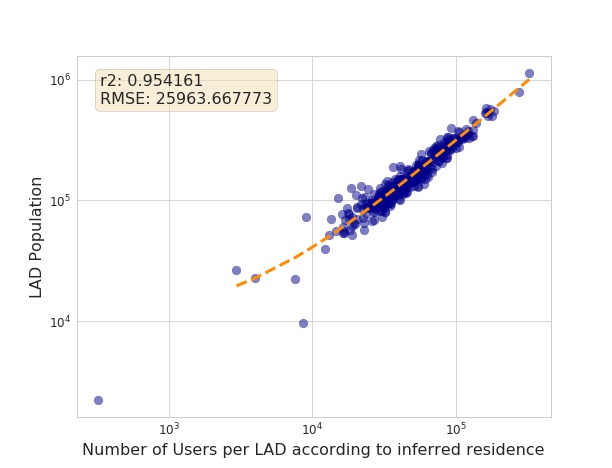}%
	\vspace{-3mm}
	\caption{Comparison between inferred residential \ac{LAD} population and the actual LAD population from census data. }
	\label{fig:home_location_validation}
    \vspace{-6mm}
\end{figure}

\paragraph*{Home Detection}
\label{sec:home_detection}
For our analysis, locating the home postcode of the end-users is important when capturing their mobility patterns. 
Home Detection algorithms are a specific kind of a wider group of algorithms used to identify meaningful places from mobility information. 
The main idea consists in using some criteria to define time slots for home, work and other activities and then use the mobility information to identify these places \cite{montoliu2010discovering, frias2010towards, isaacman2011identifying, ahas2010using}.
We estimate home location for each user at postcode granularity. 
For this, we use the cell tower to which the user connects more time during nighttime hours (12:00 PM through 8:00 AM) for at least 14 days (not necessarily consecutive) during February 2020. With that filtering, we were able to determine the home postal code for approximately 16 million users.

An inherent limitation with our inference is that the estimation of the home location distribution is influenced by the market share of the MNO, and how it reflects the general population.
In order to validate its reliability, we assigned all subjects to a Local Authority District  \ac{LAD} \cite{LAD2019} according to our home estimations, and compared with values of population estimation from the Office for National Statistics (see Figure~\ref{fig:home_location_validation}). 
The result shows a linear relationship($r^2=.955$), thus validating the representativity of the dataset. The values obtained are in line with the literature \cite{phithakkitnukoon2012socio}.


\subsection{Network Performance Statistics}
\label{sec:performance_metrics}

Using the general signalling dataset, we evaluate the average time the users spend connected to the different \ac{RAT} cells. 
We find that 4G is the most popular \ac{RAT}, with users spending on average 75\% of the time per day connected to 4G cells. 
Thus, for the network performance statistics, we focus on 4G cells as they have the highest load out of the three \acp{RAT}. 

Based on the Radio Network Performance data feed, we generate network performance statistics at the 4G radio cell level.
For each cell, we separate the following hourly performance metrics: 
the \ac{UL} and the \ac{DL} data volume (the sum of all data transferred on all cell bearers corresponding to \ac{QCI} from 1 to 8 in each direction, \ac{UL} and \ac{DL}), 
average number of active \ac{DL} users (users with active data transmission in the \ac{DL} buffer), 
average radio load (as \ac{TTI} utilization, representing the number of active \acp{UE} the LTE scheduler assigns per TTI), 
average user \ac{DL} throughput (as the average throughput over all users active in the cell in one hour, considering all bearers corresponding to \ac{QCI} from 1 to 8), 
and time (number of seconds) with active data per cell. 
We also extract hourly metrics per cell specifically for conversational voice (separating only the bearers corresponding to \ac{QCI} value 1), namely: 
voice traffic volume (total traffic with QCI equal to 1), 
average number of simultaneous voice active users, and 
the \ac{UL} and \ac{DL} average packet loss error rates. 

For all the  hourly metrics, we further aggregate them per day and extract the (hourly) median value per cell. 
This allows to capture one single value per metric per day, enabling further analysis with the daily mobility metrics 
For each of the radio cell, we attach the location metadata information from the Radio Network Topology data feed. 
We further merge this (at the postcode level) with the UK Administrative and Geodemographic Datasets to append extra information such as the geodemographic cluster for each radio cell.

\section{Mobility}
\label{sec:mobility}

In this section we present how the evolution of the pandemic and the social distancing measures impacted  mobility by analyzing the change of the metrics detailed in Section~\ref{sec:mobility_metrics}.
We capture the mobility metrics of users for 10 weeks (from week 10 to week 19 of 2020), which includes time before the SARS-CoV-2 pandemic was declared in the UK on March 11th 2020 (week 11), as well as during the government imposed  measures to tackle the emergency. 
The total number of end-users whose data we aggregate for this study is approximately 16 million, and, unless otherwise specified, for all metrics we report for every day the percentage of change in the average daily value compared to average weekly value in week 9 (23 February - 1 March 2020).

\subsection{National Mobility}

\begin{figure}[t]
	\centering
	\begin{subfigure}{0.99\columnwidth}
		\includegraphics[width=\linewidth]{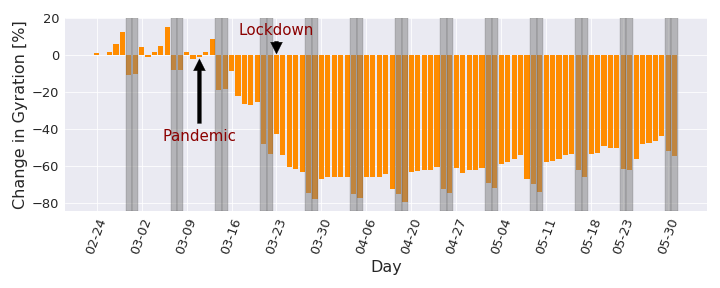}%
		\vspace{-3mm}
		\caption{\tiny Average gyration variation per user per day.}%
		\label{fig:UK_gyration}%
	\end{subfigure}%

	\begin{subfigure}{0.99\columnwidth}
		\includegraphics[width=\linewidth]{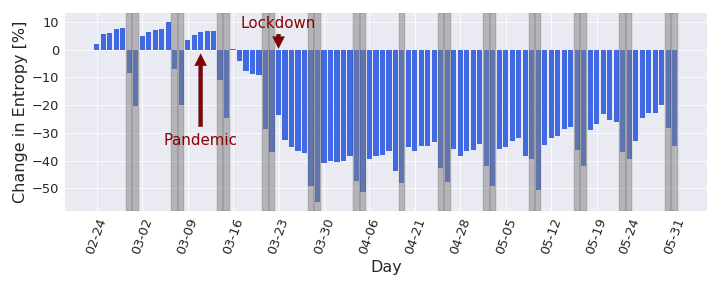}%
		\vspace{-3mm}
		\caption{\tiny Average entropy variation per user per day.}%
		\label{fig:UK_entropy}%
	\end{subfigure}%
\vspace{-3mm}
\caption{\small Percentage of change in the average value per user for radius of gyration and entropy, compared to their average value in week 9. Shaded bars correspond to weekends.  }
	\label{fig:UK_mobility}
	\vspace{-3mm}
\end{figure}

We start our analysis by investigating the nation-wide time series for radius of gyration and the mobility entropy in Fig~\ref{fig:UK_mobility}.

The average gyration evolution (Figure~\ref{fig:UK_gyration}) shows the reduction in the total area that users cover in their daily routines in reference to the average value over week 9\footnote{Note that during the weekdays of week 9, the gyration is larger while during the weekend is smaller, yielding the aforementioned average.}. We note that people started implementing social distancing recommendations even before lockdown was enforced, with a decrease of 20\% in gyration in week 12.
With the government imposing the nation-wide lockdown in week 13, we also observe a steep decrease in gyration, with a drop of 50\% towards the end of week 13 compared to the usual value from week 9. 
Mobility entropy per user follows a similar trend (Figure~\ref{fig:UK_entropy}). All metrics show a steep decrease in weeks 13-14, following the "stay-at-home" being enforced. 
In the following weeks, we note a slight relaxation, with mobility marginally increasing. It is worth noticing that, the reduction of entropy is smaller than the reduction of gyration. This indicates that people, besides moving significantly less, tend to move close to their home location.

\begin{figure}[t] 
	\centering
		\includegraphics[width=.95\linewidth]{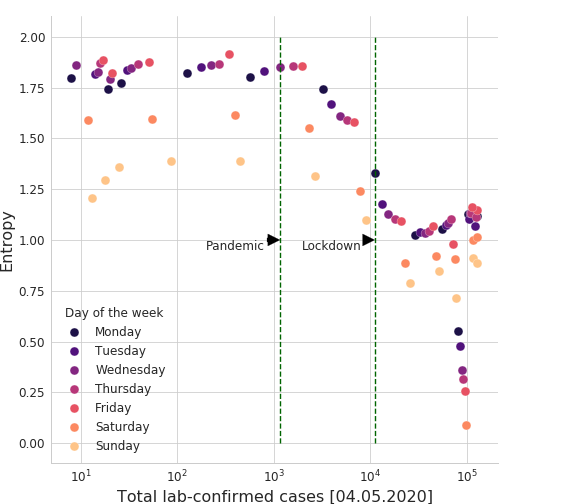}%
		\vspace{-3mm}
		\caption{\small Entropy variation (from week 9 to week 18) vs  cumulative number SARS-CoV-2 infections per day. Each point represents a different day. Colors encode the different days of the week (yellow shows week-end). } 
	\label{fig:scatter_entropy_covid}
	\vspace{-5mm}
\end{figure}

Figure~\ref{fig:scatter_entropy_covid} captures the correlation between the average mobility entropy per user and the nation-wide cumulative number of lab-confirmed SARS-CoV-2 cases, as reported by Public Health England\cite{GOV1}.
Each point in the scatterplot represent a different day; we capture the interval between February 23rd until May 4th, 2020. We note that the mobility reduction is not impacted by the number of reported cases, i.e., there is not a correlation between number of cases and mobility, but rather mobility is impacted by public announcement and lockdown measures. The decrease in the entropy starts only after the pandemic is declared (vertical red line in Figure~\ref{fig:scatter_entropy_covid}, coinciding with 1,000 confirmed cases), and becomes significant after the  lockdown. 
\noindent

\textbf{Takeaway:} Mobility metrics (gyration and entropy) show a steep decrease in people's mobility in weeks 13-14, following the "stay-at-home" order being enforced. 
We find no correlation between this reduction in mobility and the number of confimed COVID-19 cases, showing that only the enforced government order was effective in significantly reducing mobility. We also notice mobility slightly increases from week 15 despite the lockdown still being enforced.

\subsection{Regional Mobility}
\begin{figure}[!t]
	\centering
	
	\begin{subfigure}{\columnwidth}
		\includegraphics[width=\linewidth]{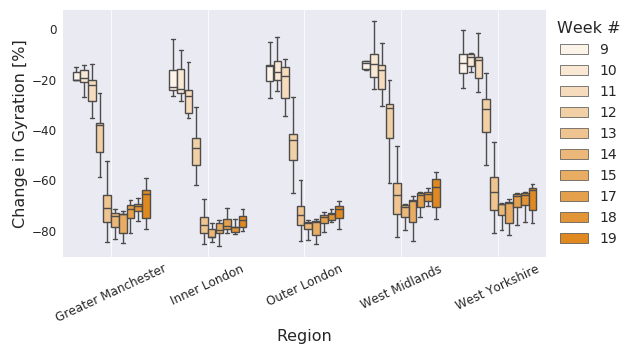}%
		\vspace{-3mm}
		\caption{\tiny Gyration.}%
		\label{fig:region_gyration}%
	\end{subfigure}%
	
	\begin{subfigure}{\columnwidth}
		\includegraphics[width=\linewidth]{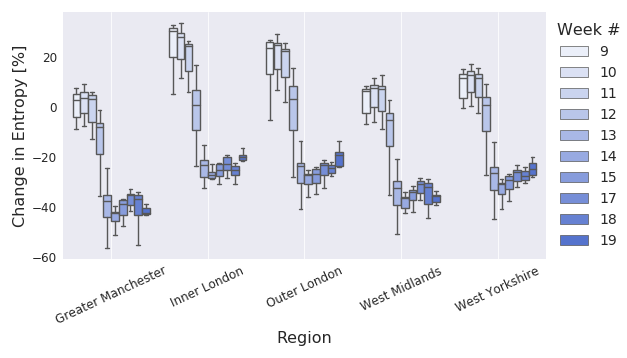}%
		\vspace{-3mm}
		\caption{\tiny Mobility entropy.}%
		\label{fig:region_entropy}%
	\end{subfigure}%
	
	\vspace{-5mm}
	\caption{\small Variation in the average gyration and entropy per region, compared to the national average during week 9.}
	\label{fig:region_mobility}
	\vspace{-6mm}
\end{figure}

We now focus our analysis on five different regions in order to observe potential geo-spatial difference in the mobility pattern changes.
We select the regions that are best represented in our dataset with more than 500,000 users, namely Inner London (700k users), Outer London (1,1 million users), Greater Manchester (700k users), West Midland (600k users) and West Yorkshire (500k users).  
For each region, we capture the variation of the two mobility metrics (Figure~\ref{fig:region_mobility}) in reference to the nation-wide average value of the metric. 
The evolution of the metrics shows clearly the impact of the stay-at-home measures in every region, with a sharp decrease in weeks 13-14 in the values of all metrics. We note that for London (both Inner and Outer London), reference values for gyration are below national average (20\% below the average for each corresponding week-- see Figure~\ref{fig:region_gyration}), while the reference value for mobility entropy is higher (20\% above the average, see Figure~\ref{fig:region_entropy}). This shows that within London, in general, people move more randomly and with less predictable mobility pattern, but cover smaller areas than the national average. 
This analysis also bring to our attention the regional differences in how people relax the mobility restrictions, with an increase in mobility in London and West Yorkshire in weeks 18-19. This is not the case for the regions of Greater Manchester and West Midlands, where mobility is consistently low after week 13.
Finally, metrics distributions have little variance in all regions, and all percentiles are close to the median, following similar trends.

\textbf{Takeaway:} The impact of the lockdown is consistent over different regions in the UK, showing that people respected the lockdown, regardless where they live. We do, however, find regional differences in how people relax the mobility restrictions, with an increase in mobility in London and West Yorkshire in weeks 18-19. This is not the case for the regions of Greater Manchester and West Midlands, where mobility is consistently low after week 13.

\subsection{Geodemographic Mobility}
To capture the changes in users' mobility in correlation with different social and demographic characteristics of their residence area, we study the variation of mobility metrics across different geo-spacial clusters as defined by the UK Office for National Statistics (ONS)\cite{ons1}.
This consists of eight groups meant to be illustrative of the characteristics of areas in terms of their demographic structure, household composition, housing, socio-economic characteristics and employment patterns.
Each of the eight categories provides the most generic descriptions of the corresponding population group in the UK (Table \ref{tab:socio_economic_groups}).

\begin{table*}[t]
\centering
\caption{Geodemographic clusters (2011 OAC).}
\vspace{-3mm}
\begin{tabular}{|c|p{12.5cm}|}
\hline
   \textbf{Name}                     &                 \textbf{Definition}                     \\
   \hline \hline
Rural Residents             & Rural areas, low density, older and educated population \\
\hline
Cosmopolitans               & Densely populated urban areas, high ethnic integration, young adults and students                                                               \\
\hline
Ethnicity Central           & Denser central areas of London, non-white ethnic groups, young adults                                                                       \\
\hline
Multicultural Metropolitans & Urban areas in transition between centres and suburbia, high ethnic mix         \\
\hline
Urbanites                  & Urban areas mainly in southern England, average ethnic mix, low unemployment\\
\hline
Suburbanites                & Population above retirement age and parents with school age children, low unemployment                    \\
\hline
Constrained City Dwellers   & Densely populated areas, single/divorced population, higher level of unemployment                                 \\
\hline
Hard-pressed Living         & Urban surroundings (northern England/southern Wales), higher rates of unemployment    \\
\hline
\end{tabular}
\label{tab:socio_economic_groups}
\vspace{-4mm}
\end{table*}

\begin{figure}[t]
	\centering
	
	\begin{subfigure}{\columnwidth}
		\includegraphics[width=\linewidth]{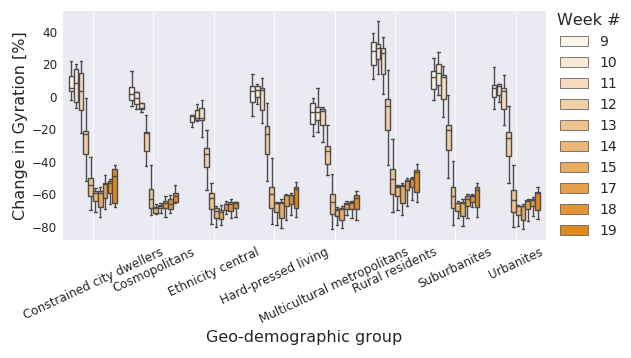}%
		\vspace{-3mm}
		\caption{\tiny Gyration.}%
		\label{fig:socio-economic_gyration}%
	\end{subfigure}%
	
	\begin{subfigure}{\columnwidth}
		\includegraphics[width=\linewidth]{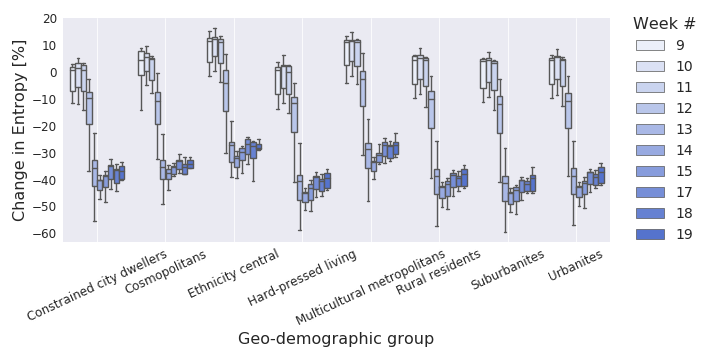}%
		\vspace{-3mm}
		\caption{\tiny Mobility entropy.}%
		\label{fig:socio-economic_entropy}%
	\end{subfigure}%
\vspace{-4mm}
		\caption{\small Variation in average gyration and entropy per geodemographic cluster, compared to the national average in week 9.}
	\label{fig:socio-economic_mobility}
	\vspace{-8mm}
\end{figure}

\begin{figure*}[h] 
	\centering
		\includegraphics[width=.99\linewidth]{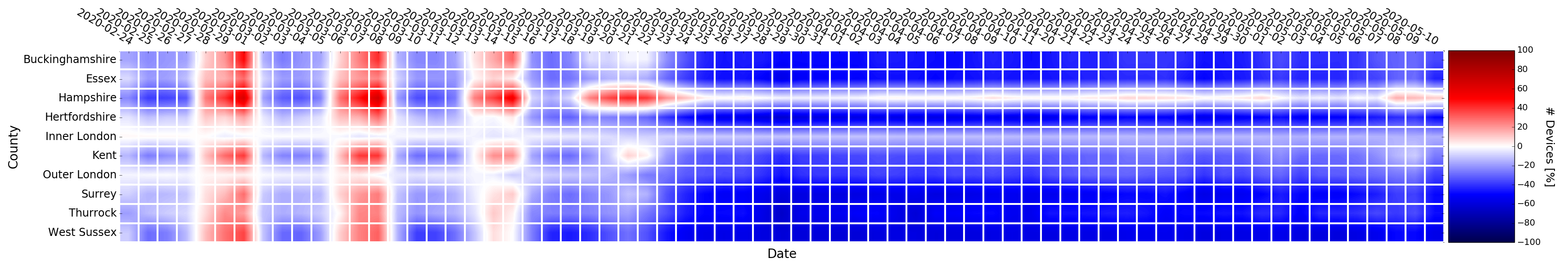}%
		\vspace{-3mm}
	\caption{\small Mobility matrix of users with residence in Inner London towards other counties in the UK. Each cell shows the variation per day (on the x axis) in the number of resident from Inner London active in the county marked in the y axis compared to the median value over week 9. We include in our mobility matrix the top 10 counties in terms of receiving inbound residents from Inner London according to the average in week 9. } 
	\label{fig:MM_InnerLondon_home}
	\vspace{-5mm}
\end{figure*} 

We break down the mobility metrics variations per week in each of the above-mentioned area groups (Figure~\ref{fig:socio-economic_mobility}). 
We find that mobility in rural areas is normally higher than the nation average and people usually cover wider areas in their daily movement (weeks 9-11 in Figure~\ref{fig:socio-economic_gyration}). 
Contrariwise, in highly populated urban areas (such as Cosmopolitans or Ethnicity central), the population covers smaller areas (weeks 9-11 in Figure~\ref{fig:socio-economic_gyration}), but the predictability of their mobility patterns is lower (higher entropy in weeks 9-11 in Figure~\ref{fig:socio-economic_entropy}). 
All the different socio-economic groups present the same significant drop in mobility, with a transition period in week 12 and a steep drop from week 13. 
Gyration decreases across all groups by more than 50\% of the national average in week 9.  As it was observed in the previous section, the ethnicity central group (which would basically correspond to Inner London), the reduction of gyration is the highest of all the groups but their reduction in entropy is the smallest, implying that they have reduced they area of mobility the most, but within that area their movement is more random. This is probably related to the high density of commercial services in central London. Finally, we observe that metrics' distribution has little variance in all geodemographic clusters.

\textbf{Takeaway:} All users in different geodemographic clusters present the same significant drop in mobility, with a transition period in week 12 and a steep drop from week 13.

\subsection{Temporary Relocation from London}

In this section, we measure how many people from London decided to move elsewhere during (part of) the analyzed time frame. 
To evaluate the temporary relocation of Inner London residents to secondary locations, we generate its mobility matrix at a county level (see Figure~\ref{fig:MM_InnerLondon_home}). 
For each Inner London resident (obtained using the method described in \ref{sec:mobility_metrics}), we check the top 20 locations (at county level) that they visit during each day (cf. Section~\ref{sec:mobility_metrics}). If none of the visited locations during a day matches their home county (Inner London in this case) we are able to identify relocations, i.e. people with residence in London that are elsewhere.
Figure~\ref{fig:MM_InnerLondon_home} shows the variation in the number of Inner London residents who are present in the different counties (we capture a different county per row) per day, compared to the average number we identified in week 9, before the lockdown.

By looking at the Inner London line we observe a permanent 10\% decrease in the number of Inner London residents who actually are present in their area of residence from week 13 onward (after the lockdown is imposed).
Contributions to the decrease in the number of Inner London residents who remain in their inferred homes during the lockdown period may include, for example, students who left campuses in London after schools closing on the 19th of March, or long-term tourist leaving Inner London center areas, or London residents who decided to spend the lockdown in their second residences. 
In particular, we observe an increase in the number of people from London who relocated to the Hampshire area during most of the duration of the lockdown. 

Further, we note that mobility patterns of Inner London residents changed significantly after the stay-at-home order was imposed. 
Specifically, before the stay-at-home recommendations, we observe an increase of Londoners spending the weekend in other counties in the UK. This pattern disappears starting from weeks 11 and 12, concurrent with the recommendation of social distancing. 
We capture a large variation in the number of people travelling from Inner London to outside areas such as East Sussex on the 21st-22nd of March, just prior to the stay-at-home order. 
We also observe an additional increase of Londoners going to Hampshire for the weekend by the end of April (also in Kent but less so), consistent with the overall (slight) increase observed at the end of the period in the overall mobility metrics.

\textbf{Takeaway:} We find a sustained 10\% decrease in the number of Inner London residents who actually are present in their area of residence from week 13 onward (after the lockdown is imposed). We capture a large variation in the number of people travelling from Inner London to outside areas such as East Sussex on the 21st-22nd of March, just prior to the stay-at-home order being enforced.

\section{Mobile Network Performance}
\label{sec:performance}

In this section, we analyze mobile network performance indicators to investigate how changes in people's mobility induce shifts in the usual traffic patterns we observe in the radio network, and further impact the mobile network operations. 

\begin{figure}[!t]
	\centering
	\includegraphics[width=\linewidth]{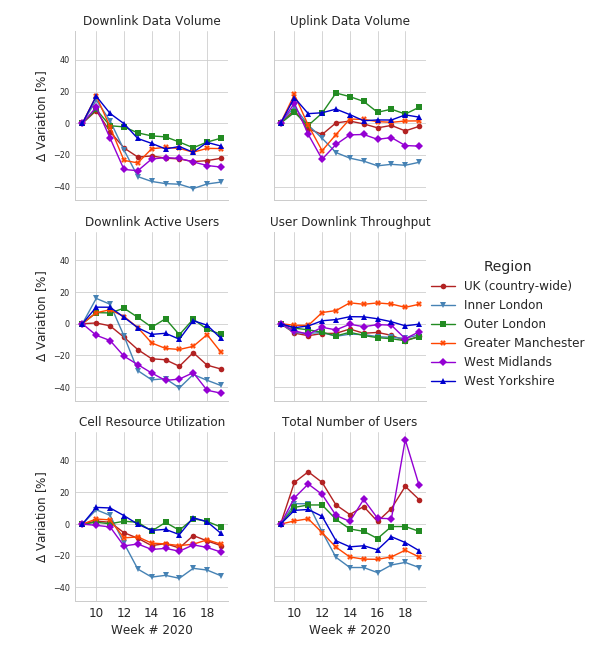}%
	\vspace{-5mm}
	\caption{\small MNO performance characterization, using metrics for all data traffic. Each plot corresponds to a metric, as we indicate in the title of the plot. We show the median values for the delta variation percentage for each metric over one week week (each point on the x-axis corresponds to a week \#), using as a baseline for comparison the median value in week 9. Each line corresponds to different geo-regions within the UK, including "UK - all regions", which accounts for the entire country.}
	\label{fig:data_UK}
	\vspace{-7mm}
\end{figure}

\subsection{Data Traffic Evolution}

We investigate the dynamics of the network performance metrics we defined in Section~\ref{sec:performance_metrics} across the UK, aggregating all bearers per cell, including voice and data traffic (i.e., we aggregate all bearers for \ac{QCI} 1 to 8). 
Our analysis spans a period of 10 weeks, from 23rd of February until the 10th of May. 
Unless otherwise stated, for all metrics we present the delta variation percentage of each metric from its median value over week 9. 

Figure~\ref{fig:data_UK}: \textit{Downlink Data Volume} captures the variation of the downlink data volume during the weeks we mark on the x-axis. 
Throughout the UK, we note an increase of 8\% in the average data volume per cell in week 10. 
This is followed by a steady decrease, with the lowest registered volume in week 17 (-24\% compared to week 9). 
Our observation is consistent with people's mobility decrease after the stay-at-home order (Section~\ref{sec:mobility}).
We conjecture that after lockdown people relied less on the cellular network for data connectivity (e.g., using home WiFi connectivity instead), thus contributing to the surge of traffic reported by residential ISPs.\footnote{See \url{https://www.cnbc.com/2020/03/27/coronavirus-can-the-internet-handle-unprecedented-surge-in-traffic.html} for instance.} 

Indeed, we observe in Figure~\ref{fig:data_UK}: \textit{Uplink Data Volume} a significant reduction of active downlink users per hour per cell compared to week 9 throughout the entire UK (all regions), with a minimum number of active users in week 19 (-28.6\% users compared to week 9). 
Note that, despite we observe a slight recovery of mobility in week 14 (cf.~Section~\ref{sec:mobility}), this is not sufficient to have an impact on the downlink data volume, the number of active users, or the radio load per cell, which do not increase.

We find a small reduction in the downlink user throughput in the 10 weeks we analyze (Figure~\ref{fig:data_UK}: \textit{User Downlink Throughput})), with the lowest value being a 10\% drop compared to week 9. This follows the same trend as the decrease in downlink traffic volume. 
This is unexpected, as in presence of less traffic (cf.~Figure~\ref{fig:data_UK}: \textit{Downlink Data Volume}) and overall lower active downlink users per cell (cf.~Figure~\ref{fig:data_UK}: \textit{Downlink Active Users}), the per user throughput should increase. 
We also observe that congestion in the radio network is not the cause of this decrease because the radio load also decreases (~Figure~\ref{fig:data_UK}: \textit{Cell Resource Utilization} shows a 15.1\% radio load decrease in week 16 compared to week 9).
We conjecture the per-user throughput is application limited and not network limited during the weeks of the analysis. 
This is in line with the choice of many content providers to reduce content quality during the pandemic and by the shift of more data-intensive applications to wifi access.~\footnote{\url{https://www.theverge.com/2020/3/20/21187930/youtube-reduces-streaming-quality-european-union-coronavirus-bandwidth-internet-traffic}}

We also investigate the variation of the total uplink data volume per cell in the UK (Figure~\ref{fig:data_UK}: \textit{Uplink Data Volume}). 
We find overall modest changes for this metric, with, less than 5\% decrease in the median daily volumes in week 18. 
This suggests that applications making intensive use of the downlink (e.g., video streaming) suffered a significantly higher traffic reduction than applications with symmetric uplink/downlink usage (e.g., audio/video conferences, or voice traffic). 
In particular, as we present in the next section, we observe that voice traffic (corresponding to \ac{QCI}=1) has significantly increased during the lockdown period.

Altogether, because the downlink data volume is one order of magnitude larger than the uplink data volume, the volume of data traffic carried by the network experienced an overall reduction of roughly 20\%, reverting to 2019 traffic volume (when the MNO had less customers and applications were less bandwidth hungry).

\ael{should we just remove this?}
Finally, we note that metrics' distribution across cells does not significantly change across weeks with respect to the reference week 9. 
A wide distribution is common in large operational environments, where cells support different \acp{RAT} types across the country (e.g, LTE, LTE-A), different number of customers each cell serves, different radio coverage, etc.
The only exception we note is in the 90th percentile in the downlink active user per hour per cell (not picture for readability reasons), which slightly reduces during the lockdown phase, thus showing the nation-wide impact on the users connectivity habits. 

\textbf{Takeaway:} We find an overall reduction (-24\% in week 17 compared to week 9) in downlink data traffic volume throughput the UK during lockdown. Unintuitively, we also note a decrease in the average user throughput. Though a decrease in radio cell resource utilization and in overall traffic may have hinted to the opposite, we find the throughput to drop by at most 10\%, likely due to application throttling. We also observe uplink traffic volume has little variation. 

\begin{figure}[!t]
	\centering
	\includegraphics[width=.9\linewidth]{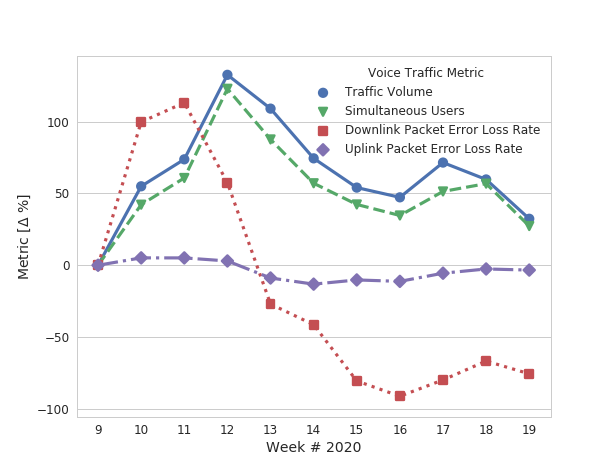}%
	\vspace{-3mm}
	\caption{\small Analysis of 4G voice traffic (corresponding to QCI=1) patterns in the UK. We show the median values  for the delta variation percentage, which we generate compared to the national average during week 9 for all the UK, over a period of 10 weeks (each point corresponds to a week \# we mark on the x-axis), from the end of February until mid-May. Each line corresponds to different metric.}
	\label{fig:voice_UK}
	\vspace{-7mm}
\end{figure}

\subsection{Voice Traffic}
\label{sec:voice}
We now analyze the changes in voice traffic patterns, to further drill down on the potential dynamics that impact the total volumes of data traffic we observe.
We filter 4G traffic marked with \ac{QCI} equal to 1 (conversational traffic), and analyze the performance metrics we previously explained in Section~\ref{sec:performance_metrics}. 
Figure~\ref{fig:voice_UK} shows the evolution of these metric over the period of 10 weeks we analyze in March - May 2020, for all the UK compared to the average of week 9. 

We note that in week 12 there is a spike of 140\% in the median value of voice traffic volume (Figure~\ref{fig:voice_UK}: \textit{Traffic Volume}) and a significant increase of its top 90 percentile value. 
This corresponds to a predicted seven years of growth in voice traffic in the operator's network, which the MNO had to accommodate in the space of few days (week 12 in Figure~\ref{fig:voice_UK}: \textit{Traffic Volume}). Though striking, we find that the increase in voice traffic is not enough to compensate the decrease in other traffic classes, resulting in the overall decrease in traffic presented in the previous section.
The surge in voice traffic volume is consistent with a spike in the average number of simultaneous voice users per cell (Figure~\ref{fig:voice_UK}: \textit{Simultaneous Users}).
This aligns with the time when the pandemic was declared (11th of March), and the lockdown restrictions started in the UK (23rd of March, week 13). 
This is an anticipated result of the lockdown and the subsequent changes in people's mobility, since more are likely to carry on voice conversations in these circumstances. 

We report in Figure~\ref{fig:voice_UK}: \textit{Uplink Packet Error Loss Rate} and Figure~\ref{fig:voice_UK}: \textit{Downlink Packet Error Loss Rate} the daily average uplink and downlink packet loss error rate per cell respectively. While the uplink packet loss decreases during the pandemic period, we observe a significant increase in the downlink packet loss during weeks 10-11-12, after which it decreases below normal values. This is again unexpected, as the radio network is less congested and there is less downlink traffic (see previous section). The issue during these weeks was, in fact, that the unexpected surge in the amount of voice traffic exceeded the capacity of the inter-MNO interconnection infrastructure. 
The rapid response of the network operators and service providers quickly restored the DL error below the normal values.

\textbf{Takeaway:} We observe a spike of 140\% in the median value of voice traffic volume around lockdown. Although the MNO was dimensioned to accommodate the traffic, this unexpected surge exceeded the capacity of the interconnection infrastructure between MNOs. The rapid response of the network operators and service providers restored the network status back to normal quickly after week 13.

\subsection{High Density Areas}

To capture regional traffic dynamics, we now zoom into the network performance of five counties with high density of users in our dataset. 
Specifically, we analyse the network performance in Outer London, Inner London, Greater Manchester, West Midlands and West Yorkshire. 
\ael{the main motivation here is the fact that the operator uses different vendors in different regions, but we cannot say which and where in the paper.}

We analyze four performance metrics, namely the daily uplink/downlink data traffic volume per cell, the number of active downlink users per cell, and the downlink average user throughput. We compleent this analysis by also showing the total number of users connected (both active and idle), and the cell resource utilization. 
We report on the delta variation percentages of each metric per week with respect to the nation-wide average for each metric observed in week 9 (Figure~\ref{fig:data_UK}, plot lines marked with the region names we mention above). We capture in our analysis 10 weeks, from the 23rd of February until the 10th of May 2020. 
We observe that while the overall trends in the five counties are similar to those over the entire UK (Figure~\ref{fig:data_UK}, plot lines marked \textit{UK}), the intensity of the different trends varies significantly from one region to another.

Regarding the downlink data traffic volume (Figure~\ref{fig:data_UK}: \textit{Downlink Data Volume}), we observe in all regions a mild increase in week 10 (ranging between 9\% and 17\% depending on the region), followed by a decrease in the following weeks reaching the minimum in weeks 17 and 18. However, the decrease is much larger in Inner London (with a 41\% decrease) than in the other regions, which decrease by about 20\%. 
Further, in outer London, there is the smallest decrease, with only 15\%. 

We also see in Figure~\ref{fig:data_UK}: \textit{Downlink Active Users} that during lockdown the number of active downlink users per cell significantly decreases in Inner London (e.g., -40\% in week 15), while it is almost constant and even slightly higher in Outer London. 
These different trends between commercial/business and residential areas are also visible for other regions as well (i.e., West Midlands/Greater Manchester more business compared to Yorkshire more residential). 

An additional contributing factor to the decrease of traffic and users in Inner London is the temporary relocation of people living in Inner London to other counties outside the city. In Section~\ref{sec:mobility} we saw that approximately 10\% of inferred residents were not signalling against radio cells covering London after lockdonw, thus hinting a potential relocation and partially explaining the reduction in the data traffic volume. 
We make a detailed analysis of the different districts within Inner London in Section~\ref{sec:london}. 

Regarding the uplink data volume (Figure~\ref{fig:data_UK}: \textit{Uplink Data Volume}), we observe different trends across regions. 
The uplink traffic volume grows for all regions in week 10. 
This growth is followed by a decrease in Inner London and West Midlands, and by a slight increase or steadier behavior in Outer London and West Yorkshire.
The highest contrast is again in Inner/Outer London. In week 14, in Inner London, the UL data traffic volume decreased -22\%; while in the Outer London region it increased by 17\%. In the other regions, the effects are similar, but with a lower intensity.  

We conjecture these differences correlate with the different geodemographic dynamics within these regions. 
For example, Inner London includes many business, touristic, commercial and recreational areas, which remained largely empty during lockdown, potentially explaining the steep decrease in the downlink data volume in the area. 
To capture these correlations, in the following section we further study the network performance on areas with the same geodemographic profiles (Table~\ref{tab:socio_economic_groups}). Thus, we change from the current perspective of network performance within administrative boundaries, to capturing performance metrics across clusters of areas with the different geodemographic profiles.

\textbf{Takeaway:} We observe that while the overall trends in five counties are similar to those in the entire UK, their intensity varies significantly from one region to another. We conjecture these differences correlate with the different geodemographic dynamics within these regions.

\begin{figure}[!t]
	\centering
	\includegraphics[width=\linewidth]{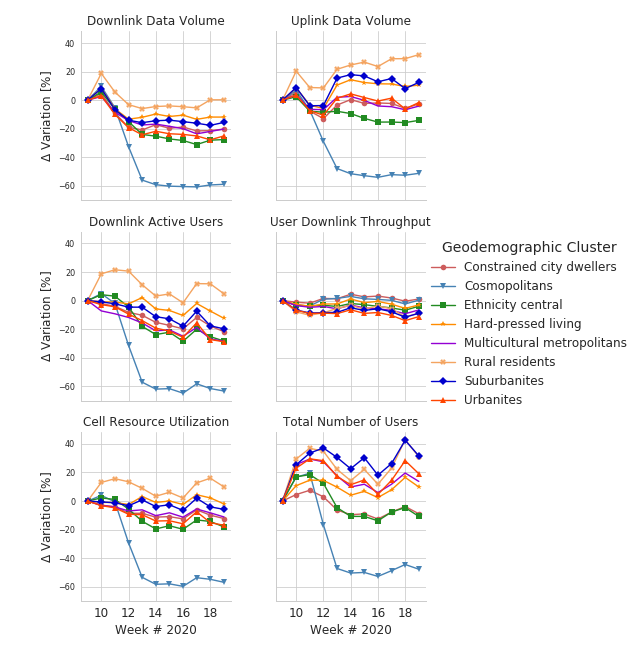}%
	\caption{\small Focused analysis of network performance metrics on geodemographic clusters (OAC 2011 dataset). 
	We show the median variation per cluster (see Table~\ref{tab:socio_economic_groups}), where each line corresponds to a different cluster.}
	\label{fig:data_portraits}
	\vspace{-5mm}
\end{figure}

\subsection{Geodemographic Clusters}

In this section, we incorporate into our analysis of network performance during  COVID-19 demographic dynamics. 
Geodemographic classifications provide summary indicators of the social, economic, demographic, and built characteristics of small areas.
To this end, we use the geodemographic clusters (Table~\ref{tab:socio_economic_groups}) that the ONS labeled to each postcode.\footnote{National Statistics Postcode Lookup (NSPL) for the United Kingdom as at February 2020. \url{https://geoportal.statistics.gov.uk/datasets/national-statistics-postcode-lookup-february-2020}}
We group the postcode areas in the corresponding geodemographic clusters and extract the network performance statistics (Section~\ref{sec:performance_metrics}) per cluster. 
This analysis allows us to understand  
potential network performance differences between groups of areas with distinct mobility patterns across the entire UK (e.g., rural residents areas vs. cosmopolitan areas, see Figure~\ref{fig:socio-economic_mobility}), as we already observed different behavior for different areas in the previous section.

We find that the behaviour for most areas, with the exception of "Rural residents" and "Cosmopolitan" areas, exhibits similar trends than the ones observed in UK nation wide, with a decrease in downlink traffic and a stable uplink traffic volume. 
In areas within the "Rural residents" cluster, the average volume of downlink data traffic per cell remains largely stable after the lockdown period.

Figure~\ref{fig:data_portraits} shows that the average volume of downlink data traffic per cell in areas with majority of "Cosmopolitan" residents decreased dramatically after week 13, coinciding with the stay-at-home order. 
The areas within the "Cosmopolitans" cluster correspond to densely populated urban areas, with high ethnic integration, where young adults and students reside (see Table~\ref{tab:socio_economic_groups}). 
These areas include the neuralgic centers for business, commercial, recreational, educational activities throughout the country.
The pattern we observe is different from the rest of the areas. 
This is likely to be caused by a dramatic reduction of people in these places, since are areas with a lower number of residents but with a lot of people visiting them due to work and recreational purposes. 
This is consistent with the observations we made in the previous section for Inner London, where approximately 45\% of postcode areas cluster within the "Cosmopolitans" geodemographic group, while 50\% of postcode cluster in the "Ethnicity Central" group. 

To further verify this correlation, in Figure~\ref{fig:data_portraits} we show the total number of users connected to the network, and observe a sharp decrease of up to -50\% in the "Cosmopolitan" areas. 
We calculate the correlation between the total number of users (both idle and active users) and the downlink data volume for the geodemographic clusters. We find a high correlation for the "Cosmopolitan" (+0.973) and for "Ethnicity central" (+0.816) areas, low correlation for "Rural residents" areas (0.299), and a negative correlation for "Suburbanites" (-0.466), where the increase of users resulted in a decrease of downlink data volume (Figure~\ref{fig:data_UK}). In these areas, the drop in downlink data volume is likely impacted by other factors (e.g., offloading to WiFi, changes in applications used). 

\textbf{Takeaway:} 
Areas within "Rural residents" and "Cosmopolitan" geodemographic clusters deviate from the trends we observed in UK nation wide.
In the former, the average volume of downlink data traffic per cell remains largely stable after the lockdown.
Conversely, in the latter we observe a sharp decrease of up to -50\% in the total number of users connected after lockdown, which led to a dramatic decrease in the downlink data traffic volume. 

\section{London-centric Analysis}
\label{sec:london}

In this section, we focus our analysis on the London area (i.e., the Inner London county).
London is particularly interesting because of the large concentration of users in a single area, high mobility, diversity in user's profiles, and also because, unfortunately, the pandemic was particularly severe in the city.
Our goal is to dissect how the changes in the mobility patterns of the population residing in this region (approximately 700,000 users in our dataset) resulted in changes in their habits of using the mobile network service. 

\begin{figure}[!h]
	\centering
	\includegraphics[width=\linewidth]{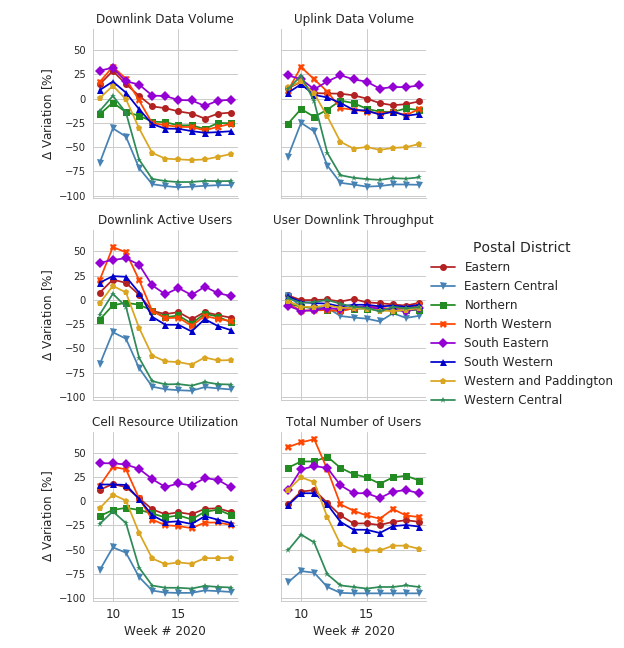}%
	\vspace{-3mm}
	\caption{\small Network performance metrics for Inner London. Each line corresponds to a Postal District within London. 
	We show the median variation percentage (on the y axis) per postal district, over a period of 10 weeks (on the x axis). }
	\label{fig:london_data}
\vspace{-4mm}
\end{figure}

\begin{figure}[!h]
	\centering
	\includegraphics[width=\linewidth]{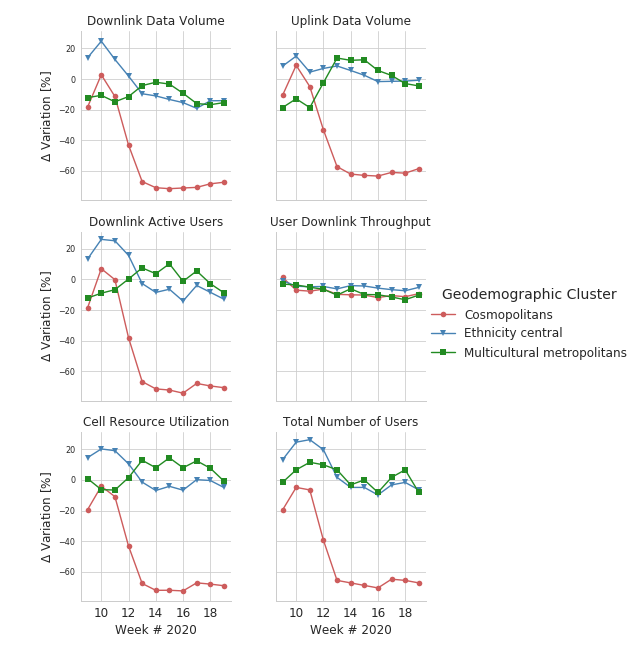}%
	\vspace{-2mm}
	\caption{\small Network performance metrics per geodemographic clusters. 
	We show the median variation percentage (on the y axis) per cluster (see Table~\ref{tab:socio_economic_groups}), where each line corresponds to a different cluster, over a period of 10 weeks (on the x axis). }
	\label{fig:london_geodemographic}
	\vspace{-5mm}
\end{figure}

\subsection{Network Performance}
Figure~\ref{fig:london_data} shows the network performance evolution based on data volume metrics, compared to median values over week 9 within London. We break down the analysis for each of the Postcode Districts within London, and we only depict the median values over each week (for readability reasons).
We note a general decrease in the total volume of downlink traffic (Figure~\ref{fig:london_data}: \textit{Downlink Data Volume}), as well as in the volume of uplink traffic (Figure~\ref{fig:london_data}: \textit{Uplink Data Volume}). 
The decrease is particularly large for Eastern Central (EC) and Western Central (WC). 
In the WC district, the uplink and the downlink traffic volumes exhibited a decrease of over 80\% between weeks 14 and 19, while the EC district experienced a similar trend, with a decrease of more than 70\% compared to week 9. 
These two districts cover the central area of London, where the density of population is also lower (i.e., approximately 30,000 residents in EC compared to 400,000 residents in SW) and includes many seasonal residents (e.g., tourists), business and commercial areas. 
The decreasing volume of data is likely due to a major reduction of people in the area, which is consistent with the decrease we observed in the average number of radio users per sector (see Figure~\ref{fig:london_data}: \textit{Total Number of Users}). 
We note a similar trend both in the percentage of cell resources occupancy (Figure~\ref{fig:london_data}: \textit{Cell Resource Utilization}) and in the average number of users per cell (Figure~\ref{fig:london_data}: \textit{Downlink Active Users}).  

We find that the trends in the Northern (N) district detach from the rest of the districts.
In particular, we note an increase in the number of downlink users (varying between 10\% and 23\% in weeks 10 to 14), while the volume of downlink traffic keeps stable (unlike the other postcodes, where we note a decrease). This hints towards potential hot spots in the mobile network moving within London from central areas (EC, WC) to the north (N). 

\textbf{Takeaway:} Performance in the central postal districts of London (EC and WC) differ from the rest, with decreases of over 70\% in the downlink data traffic. This is due to a major reduction of people in the area, which is consistent with the decrease we observed in the average number of total radio users in those sectors. 

\subsection{Geodemographic Clusters}

We further dissect the performance in London in relation to the geodemographic clusters. 
We find that only three clusters map to the area of London, each with different patterns. 
Figure~\ref{fig:london_geodemographic} shows the evolution of four performance metric over the period of analysis (median values of delta variation percentage per week compared to corresponding median values in week 9 in London). 
We find again a sharper decrease in the areas marked as "Cosmopolitans", largely matching the behaviour we observed in the EC and WC districts, due to the severe reduction of users in these areas.
We find that all areas follow the same trends for the user downlink throughput, which is consistent with the UK-wide observation (Figure~\ref{fig:data_UK}: \textit{User Downlink Throughput}). 
However, for the uplink/downlink traffic volume, we note distinct patterns for the three clusters, showing distinct usage patterns across the different areas. 
We observe a sharp decrease in the traffic volumes in the radio cells in Cosmopolitans cluster areas (decrease of more than 50\% in week 13 in the uplink and the downlink data volume), while cells in the "Multicultural clusters" cluster areas show an increase the mobile traffic volume (e.g., 40\% increase in the uplink data volume). 
These correlate with the patterns we observe for the number of active downlink users in the same areas (i.e., a more than 20\% increase in week 13). 

\textbf{Takeaway:} We find again a sharper decrease in the uplink/downlink data volume in the areas that cluster as "Cosmopolitans", largely matching the behaviour we observed in the EC and WC districts, due to the severe reduction of users in these areas. Cells in the "Multicultural clusters" cluster areas show an increase the mobile traffic volume largely due to an increase in the number of active users. 

\section{Related Work}

Even though the spread of SARS-CoV-2 started only a few months ago, there are already in the literature different studies that present efforts for understanding human mobility during the interventions. Apart from  cell phone traces, as used in our study and in others \cite{jia2020population}, there are other datasets that other studies investigate. Google has available anonymized and aggregated private counts of visits to places in different categories \cite{wellenius2020impacts}\cite{google1}\cite{kraemer2020mapping}; also, Facebook trough its Disease Prevention maps has provided aggregated data of spatial distribution of individuals \cite{maas2019facebook}\cite{galeazzi2020human}; Cuebiq has provided GDPR-compliant information to model mobility in Boston~\cite{aleta2020modeling} and Italy~\cite{pepe}; and Baidu has provided in-flow and out-flow indexes for Chinese cities~\cite{chinazzi2020effect}. In general, all the previous studies highlight the drastic changes in mobility and evaluate the impact of the interventions in the spread of the virus. In the particular case of the UK, TomTom~\cite{tomtom1} observed a reduction between 60\% and 80\% in the number of trips during lockdown, which is consistent with the reduction we observe from the MNO. Also, \cite{imperial1}, by combining cell phone traces and Facebook data to identify number of journeys, concluded that mobility began to decrease around one week
before lockdown was enforced, but that the sharpest drop was after the closure, similarly to our findings using different metrics.

The overall topic of the impact of COVID-19 in network traffic has also been subject to numerous reports. For example, Comcast, which provides mobile connectivity using WiFi while at home but leverages on Verizon LTE while away, has observed and increase of 39\% in wifi and a decrease of 17\% of mobile LTE traffic.
Regarding the increase of the different types of applications, there was a 215-285\% increase in VoIP and videoconferencing traffic, 30-40\% in VPN traffic and a 20-40\% in streaming and web video consumption.\cite{comcast1}. In the context of network traffic and distance learning, of critical relevance during COVID-19, the work by Favale at al.~\cite{favale2020campus} found that incoming traffic drastically decreased, while outgoing traffic more than doubled to support online learning.
By combining datasets from one ISP, three IXPs, and one metropolitan educational network, Feldmann et. al.~\cite{feldmann2020lockdown} show that the fixed internet infrastructure was able to sustain the 15-20\% increase in traffic that happened rapidly during a short window of one week. 
Taking the point of view from Facebook's edge, Böttger et.al.~\cite{bottger2020facebook} show that that different regions of the world saw different magnitudes of impact, with predominantly less developed regions exhibiting larger performance degradations.
The Cellular Telecommunications Industry Association (CTIA) provides daily reports on the variations of voice and data minutes observed by AT\&T, T-Mobile, U.S. Cellular and Verizon. With respect to the period between the 23 of February  and the 16 of March, it was observed a surge of the upper range of voice minutes used of up to +24\% in the last week of march (week 13) and an increase of about +20\% in the upper range of data minutes\cite{ctia1}. While the surge reported in voice traffic is within the ranges observed in this paper, we do not observe the surge in data traffic reported by the main MNOs in the U.S. This may be due to different prevalence of flat fee contracts in the U.S. operators and the U.K. MNO we study. Similarly, BT in the U.K., on March the 20th 2020 \cite{bt1}, reported a decrease of 5\% in mobile data traffic which is consistent with the decrease described in this paper for week 12. It was also observed a surge in voice traffic, and as a result BT encouraged users to use landline services, especially for long conference calls.

\section{Conclusions}
\label{sec:conclusions}
The lockdown measures implemented by the UK government to limit the spread of SARS-CoV-2 virus drastically changed user mobility patterns and network traffic.
The results of this paper present an analysis of the changes in mobility and their impact on the cellular network traffic. 
Regarding mobility, we observed an overall decrease of 50\%, with non-uniform changes across different geographical areas and social backgrounds, which confirm previous findings~\cite{tomtom1}\cite{imperial1}. 
These variations translated into a surge of voice traffic (up to 150\%) accompanied by an overall decrease of download traffic (20\%), especially in densely populate urban areas (-60\%), and an increase of uplink traffic in suburbanities (10\%).
Despite significant pattern changes, the MNO was able to provide service maintaining quality standards: the radio load was below common values and per user throughput was likely application limited. We identified one issue in voice traffic packet loss due to exceeded capacity in the interconnection infrastructure MNOs use to exchange voice traffic, which was was fixed by network operation teams.

\appendix
\section{Ethical considerations}
The data collection and retention at network middle-boxes and elements are in accordance with the terms and conditions of the MNO and the local regulations. All datasets used in this work are covered by NDAs prohibiting any re-sharing with 3rd parties even for research purposes. Further, raw data has been reviewed and validated by the operator with respect to GPDR compliance (e.g., no identifier can be associated to person), and data processing only extracts aggregated user information at postcode level.
No personal and/or contract information was available for this study and none of the authors of this paper participated in the extraction and/or encryption of the raw data. 

\section{Acknowledgements}
We thank the IMC anonymous reviewers and our shepherd, Anja Feldmann, for their helpful comments and guidance.
The work of Andra Lutu was supported by the EC H2020 Marie Curie Individual Fellowship 841315 (DICE).

\newpage
\bibliographystyle{ACM-Reference-Format}
\bibliography{references}

\end{document}